\documentclass[runningheads]{llncs}
\usepackage[T1]{fontenc}
\usepackage{cite, url}
\usepackage{hyperref}
\usepackage{breakurl}
\usepackage{amsmath,amssymb,amsfonts}
\usepackage{algorithm}
\usepackage{algorithmic}
\usepackage{longtable}
\usepackage{array}
 \usepackage{booktabs} 
\usepackage{graphicx}
\usepackage{textcomp}
\usepackage{xcolor}
\begin{document}
\title{SeqShield: A Behavioral Analysis Approach to Uncover Rootkits}
\author{Paras Ghodeshwar\inst{1}\orcidID{0009-0009-2301-007X} \and
Sandeep K Shukla \inst{2} \orcidID{0000-0001-5525-7426} \and
Anand Handa\inst{3}\orcidID{0000-0003-0075-1165} \and
Nitesh Kumar\inst{3}\orcidID{0000-0003-0998-0925} \and
 }
\authorrunning{P. Ghodeshwar et al.}
% \institute{Departmet   \\
% Indinna  \\
% \email{}
\institute{Department of Computer Science and Enginerring, \\Indian Institute of Technology, Kanpur, India \\
\email{parasg@cse.iitk.ac.in}\\ \and
International Institute of Information Technology, Hyderabad, India \\
\email{sandeeps@iiit.ac.in}\\ \and
C3iHub Center, Indian Institute of Technology, Kanpur, India\\
\email{\{anand, niteshkr\}@c3ihub.iitk.ac.in}}
\maketitle              % typeset the header of the contribution
\begin{abstract}
Rootkits are among the most elusive types of malware, capable of bypassing traditional static analysis methods due to their metamorphic behavior. \textit{Signature-based} detection techniques struggle against these threats, necessitating a shift toward dynamic analysis approaches. We propose SeqShield, a behavior-based rootkit detection approach designed specifically for the Windows OS, leveraging API call sequences for dynamic behavior analysis. Instead of relying on static signatures, SeqShield examines the execution patterns of \textit{API calls}, which inherently reflect malicious intent. Analyzing API sequences, we can effectively identify rootkit-like behavior. We also employed a \textit{metamorphic code engine} to generate 10× mutated variants of rootkits, demonstrating their obfuscation strategies. SeqShield applies \textit{n-gram} analysis to extract bigram and trigram features from these API call sequences, enabling effective detection of rootkit-like activity. Among the models tested, Random Forest achieves the highest accuracy of 97.27\% (bigram) and 96.17\% (trigram). To optimize performance and decrease the dimension, we apply feature importance ranking using the Gini Impurity Index, iteratively selecting the most significant features. The optimized lower-dimensional feature matrix significantly enhances detection efficiency without sacrificing accuracy. Using the optimized feature set, our approach achieves 96.72\%  accuracy for bigrams and 97.81\% accuracy for trigrams.

\keywords{Kernel-Level Rootkits\and Rootkits\and Cybersecurity\and NGrams\and \\ 
NLP\and Gini Index\and MetaMe.}
\end{abstract}
\section{Introduction}

Malware has become a serious problem in today's digital world, and rootkits are one of the most dangerous types. They work secretly, hiding deep inside the operating system to avoid being noticed and keeping control over the computer for long periods. Their ability to compromise system integrity and security makes detecting rootkits a critical area of cybersecurity research.

Adversaries often employ rootkits to stealthily conceal the presence of malicious programs, files, network connections, services, drivers, and other critical system components. Rootkits achieve this by intercepting, hooking, and manipulating operating system API calls to obscure system information from security tools and administrators. Rootkits or rootkit-enabling functionalities can operate at various levels of a system. They may reside in user space, within the kernel, or even deeper in the system architecture, including the hypervisor, Master Boot Record (MBR), or system firmware \cite{mitreT1014}.

Kernel-level rootkits are considered the most dangerous and sophisticated type within the rootkit family, often referred to as next-generation rootkits due to their advanced capabilities, including kernel hooks and data structure manipulation. Unlike first-generation rootkits, which primarily operated in user mode, kernel-level rootkits of the second and third generations transitioned to the kernel, providing attackers with deeper access and greater control over the operating system. These features make kernel-level rootkits exceptionally difficult to detect and remove, posing a significant threat to system security \cite{doi:10.2352/ISSN.2470-1173.2021.3.MOBMU-034}. To conceal malicious activities or modify system behavioral techniques employed to develop kernel-level rootkits, such as IRP (I/O Request Packet) hooks, SSDT (System Service Descriptor Table) hooks, IDT (Interrupt Descriptor Table) hooks, DKOM (Direct Kernel Object Manipulation), and virtual file system hooking. Among these, DKOM is considered a third-generation rootkit technique as it directly manipulates kernel data structure\cite{9659710}.

To tackle these challenges, we propose SeqShield, a novel approach that leverages API call sequences to detect rootkit-like behavior. Our method utilizes NLP techniques, specifically bigram and trigram sequences, to analyze API call patterns for identifying malicious activities.

We did our experiment on the Windows operating system because it's widely used and often targeted by rootkits. Our dataset included a balanced mix of known rootkits and normal (benign) executables, which helped in training the model fairly. Malware writers/developers often use tricks to hide their code and avoid getting caught, we used an obfuscation tool to make our test cases more realistic. To study how the executable behaves, we looked at sequences of API calls made during execution. We created two types of feature sets from this data: bigrams (pairs of API calls) and trigrams (triplets of API calls). These patterns helped us spot suspicious behavior similar to rootkits. Although these feature sets contained both relevant and irrelevant features, we employed a novel approach using the Gini impurity index to rank and extract the most relevant ones. Instead of using a fixed threshold, we divided the sorted features into top-ranked chunks and identified the subset where the model's F1-score peaked. This allowed us to significantly reduce the dimensionality of the feature space while retaining high predictive performance. With fewer but more relevant features, our model still performed as well or even better than the full feature set. It also worked well in identifying new rootkits it hadn’t seen before, showing that the model is able to generalize and catch unknown threats.

 The key contribution of our research are as follows:
\begin{itemize}
    \item Developed a lightweight rootkit detection model using API call sequences, achieving high accuracy with minimal computational overhead.
    \item Used a Metamorphic Code Engine to simulate obfuscation, showing hash changes that evade VirusTotal detection, aligning with the Pyramid of Pain concept\cite{bianco_pyramid_of_pain_2013}.
    \item Applied Gini Impurity Index to rank and select key features, reducing feature size while maintaining model accuracy and efficiency.
    \item Validated that top-ranked features common across Decision Tree and Random Forest classifiers are most effective for rootkit detection.
    \item Relevant datasets, feature matrices, and algorithms will be shared upon request if deemed necessary.
\end{itemize}
\vspace{-5mm}
\section{Background}
\vspace{-3mm}
The concept of rootkits dates back to the early 1990s when the term \textit{rootkit} was first introduced in the context of UNIX-like operating systems\cite{wikipedia_rootkit}. A rootkit refers to a toolkit that allows attackers to gain and maintain unauthorized root-level access to a system. These tools were initially developed to modify the operating system’s core functionalities, enabling attackers to conceal their activities while maintaining control over the system.

The prominence of rootkits grew significantly in 2005 during the infamous Sony BMG copy protection rootkit scandal. The rootkit, called Extended Copy Protection (XCP), was surreptitiously installed on users' systems through a music CD. Once installed, it restricted user access and functionality, effectively acting as a Digital Rights Management (DRM) mechanism but at the cost of user privacy and system security\cite{wikipedia_rootkit, sony}. Since this incident, rootkits have become a focal point for both malware authors and cybersecurity researchers, with their functionality and sophistication evolving substantially over the years.
% \vspace{-5mm}
\subsection{Evolution of Rootkits}
% \vspace{-5mm}

The evolution of rootkits can be categorized into three distinct generations:
\vspace{-5mm}

\subsubsection{First-Generation Rootkits (User-Level Rootkits)}
Early rootkits operated at the user level and were primarily designed to execute malicious activities in the user mode of the operating system. These included tasks such as stealing passwords, collecting sensitive user information, and altering application-level functionality. While these rootkits were relatively simple, they paved the way for more advanced techniques\cite{nadim2023kernellevelrootkitdetectionprevention , doi:10.2352/ISSN.2470-1173.2021.3.MOBMU-034}.
\vspace{-5mm}

\subsubsection{Second-Generation Rootkits (Kernel-Level Rootkits)}
With advancements in malware development, attackers moved beyond user-level operations to target the kernel—the core of the operating system. Kernel-level rootkits operate in kernel mode, granting them deeper access and control over the system. These rootkits utilize system-level hooks to manipulate critical OS functionalities. Techniques such as hooking System Call Tables, I/O Request Packets (IRP), and Interrupt Descriptor Tables (IDT) became common, allowing rootkits to hide their presence and perform malicious tasks stealthily\cite{nadim2023kernellevelrootkitdetectionprevention , doi:10.2352/ISSN.2470-1173.2021.3.MOBMU-034}.
\vspace{-5mm}
\subsubsection{Third-Generation Rootkits (Kernel Data Structure Manipulation)}
The latest generation of rootkits focuses on manipulating kernel-level data structures to achieve invisibility and persistence. This category includes techniques like Direct Kernel Object Manipulation (DKOM) and Dynamic Kernel Object Hooking. Unlike traditional kernel hooks, DKOM modifies dynamic kernel data structures, making detection significantly more challenging. For instance, attackers can hide processes, kernel device drivers, and active ports by altering kernel-maintained data structures without disrupting normal system functionality\cite{nadim2023kernellevelrootkitdetectionprevention , doi:10.2352/ISSN.2470-1173.2021.3.MOBMU-034}.
\vspace{-3mm}
\section{Related Work and Motivation}

The detection of malicious software, i.e., malware, including rootkits, typically relies on two primary methodologies: \textit{static analysis} and \textit{dynamic analysis}. Each methodology has unique approaches, strengths, and limitations that, when combined, provide complementary strategies for identifying and mitigating malicious activity. Rootkit detection also benefits from these methodologies, with each offering distinct advantages in identifying rootkits, including kernel-level rootkits. Malware authors, including those developing rootkits, often employ obfuscation techniques to evade signature-based detection systems. In this research, we utilize MetaMe to mutate rootkit samples, demonstrating how obfuscation can bypass signature-based detection. 

The recent work has demonstrated the detection of the rootkits via three major factors: Hardware information-based rootkit detection Approach, Memory Forensics based rootkit detection, Behavior-based rootkit detection approaches. 
\vspace{-7mm}

\subsubsection{Hardware-Based Rootkit Detection}
Recent studies have leveraged \textit{hardware-level monitoring} to detect rootkits by analyzing control registers and hardware performance metrics. Liwei Zhou et al\cite{hardware_assisted}. introduced a \textit{hardware-assisted infrastructure} that monitors modifications to \textit{Control Register (CR3)}, identifying unauthorized changes indicative of rootkits. Their approach extracts features at the hardware level but performs \textit{off-chip analysis} in a trusted software environment using mathematical classification techniques. Similarly, Baljit Singh et al\cite{kernel_rootkit_hpc}. explored \textit{hardware performance counters (HPCs)} by creating synthetic rootkits (IRP, SSDT, DKOM) and capturing hardware-level execution data. While they successfully detected \textit{SSDT and IRP rootkits}, they failed to detect \textit{DKOM rootkits} due to their indirect manipulation of kernel data structures. However, Boyou Zhou et al\cite{hpc_malware}. highlighted that \textit{HPC-based detection suffers from high false positive rates}, making it impractical. Additionally, Joel A.\cite{phase_analysis} et al. proposed a \textit{power-based rootkit detection method}, leveraging \textit{nonlinear phase-space analysis} on \textit{ power fluctuations} to detect anomalies in execution patterns.
\vspace{-5mm}

\subsubsection{Memory Forensics-Based Rootkit Detection}
Memory forensics has proven to be an effective technique for \textit{dynamic rootkit analysis} by extracting key kernel structures from \textit{volatile memory}. Mohammad Nadim et al\cite{doi:10.2352/ISSN.2470-1173.2021.3.MOBMU-034}. demonstrated that forensic tools like \textit{Volatility} can analyze memory artifacts such as \textit{EPROCESS, ETHREAD, LDR\_DATA\_TABLE\_ENTRY, and CR0} to identify rootkits at the kernel level. Similarly, Xiao Wang et al\cite{tkrd}. utilized \textit{memory dumps from VMs} running on the \textit{OpenStack platform}, ensuring trusted acquisition from the controller node. Their approach combined \textit{memory forensic techniques with machine learning}, where extracted malicious features were used to train classifiers, achieving \textit{automated rootkit detection}. By leveraging multiple \textit{Volatility plugins}\cite{volatility_framework}, their model effectively captured suspicious memory modifications indicative of rootkits.
\vspace{-5mm}

\subsubsection{Behavior-Based Rootkit Detection}
Behavior-based detection methods focus on identifying anomalies in system activity caused by rootkits. Suresh Kumar et al\cite{dl_macfee}. proposed a \textit{deep learning-based detection framework}, where they collected \textit{hooked data} using McAfee’s \textit{Rootkit Detective} and trained a \textit{reinforcement learning model} to recognize abnormal behavior. The trained model was then enhanced using \textit{Generative Adversarial Networks (GANs)} for anomaly detection. Kruegel et al\cite{binary_analysis}. introduced a \textit{binary analysis approach}, statistically analyzing \textit{kernel module interfaces} for modifications that indicate rootkit presence. Patrick Luckett et al\cite{neuraln}. explored \textit{neural network-based rootkit detection}, utilizing \textit{system call timing analysis} to classify infected OS behavior. Their model, implemented in \textit{MATLAB’s Neural Network Toolbox}, achieved \textit{82.8\% accuracy} in classifying rootkit-infected system calls, demonstrating the effectiveness of \textit{behavioral analysis} in rootkit detection.

Previous research has explored various approaches, each with its own drawbacks. For instance, in hardware-based detection, the HPC method failed to detect DKOM rootkits and exhibited a high false positive rate, making practical implementation challenging. Memory forensics is one of the most effective techniques available; however, it is primarily suited for analyzing compromised systems and is difficult to implement efficiently in a live system environment. Additionally, detecting hooking-based methods may fail to identify DKOM rootkits, as DKOM does not rely on hooking techniques. This limitation motivated us to explore behavioral analysis as an alternative approach. Unlike traditional methods, behavioral analysis focuses on the actions and interactions of a program rather than its static properties. We were particularly interested in analyzing API call sequences, as they provide critical insights into how rootkits operate within a system. By examining API call patterns, we aim to uncover rootkit-like behavior, even in the presence of obfuscation techniques. We utilized MetaMe to mutate rootkit samples to further validate our approach, demonstrating how obfuscation can successfully evade signature-based detection. However, we can effectively detect rootkit activities regardless of their evasion techniques through behavioral analysis and API call sequence modeling.

\section{SeqShields}

SeqShield is a behavior analysis method designed to detect rootkit-like behavior, specifically targeting kernel-level rootkits in the Windows environment. The method is based on analyzing API sequences, which serve as fundamental parameters for a process to execute tasks, whether operating in user mode or kernel mode. API calls act as an essential interface between software and the operating system, enabling applications to request system resources, perform operations, and communicate with hardware. Every program, whether benign or malicious, interacts with the system through a sequence of API calls to execute its intended functionality. These functions reside in ntdll.dll in user mode but internally transition to kernel mode via a system call mechanism. Regardless of whether a rootkit operates at the kernel level or user level, it must rely on specific API call sequences to execute its malicious functionalities. Even kernel-mode rootkits, which operate with elevated privileges, typically require some form of user-mode execution to trigger their behavior, since user-mode processes are responsible for invoking system calls that eventually transition into the kernel. By analyzing these API call sequences, SeqShield provides valuable insights into a program’s behavior, enabling the distinction between legitimate and malicious activities while effectively detecting rootkit-like behavior.

To analyze API call sequences, we employed the n-grams method, a widely used technique in natural language processing (NLP) for modeling sequential data. N-grams represent contiguous sequences of n items (in this case, API calls) within a dataset \cite{malxcap}. This approach allows us to capture the immediate relationship between consecutive API calls, enabling a deeper understanding of execution patterns. The ability of n-grams to preserve sequential dependencies makes them particularly effective for behavioral analysis, as they help identify recurring patterns that indicate malicious activity. Unlike other complex models such as Long Short-Term Memory (LSTM) networks may struggle with shorter sequences and require additional mechanisms like attention mechanisms to weigh important sequence segments properly. By leveraging N-grams, we can analyze API call sequences to detect deviations from normal execution flow, making it a robust technique for identifying rootkit-like behavior. In our analysis, we specifically utilized bigram and trigram sequences, which capture the relationships between two and three consecutive API calls, respectively. This approach enables us to better understand the contextual dependencies within API sequences, improving the accuracy of our detection mechanism while maintaining computational efficiency.
\begin{figure}[]
    \centering
    \includegraphics[width=0.9\textwidth, height=8cm]{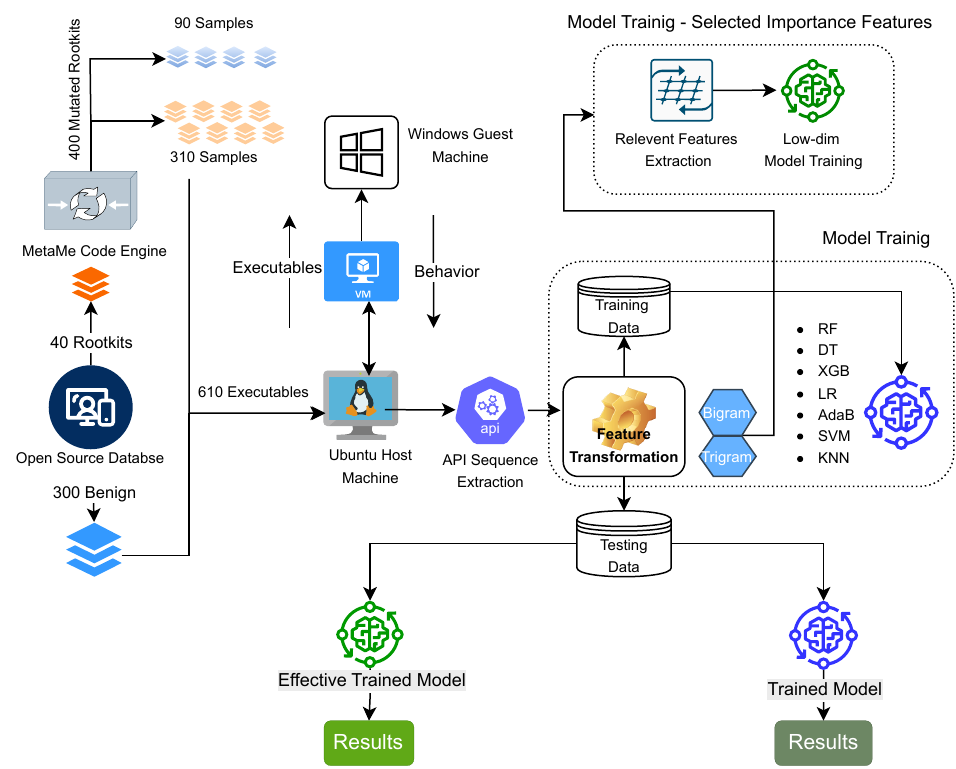}
    \caption{ Architecture of SeqShield}
    % \vspace{-7mm}
    \label{fig:architecture}
\end{figure}
\vspace{-5mm}
\subsection*{Architecture}

This section describes the architecture(Fig~\ref{fig:architecture}) of the SeqShield model, which is centered on analyzing API sequence patterns. We utilized multiple tools and modeling techniques to achieve our objectives to derive the desired results.

\subsection{Experimental Setup}

Our experiments utilize multiple tools to collect and process data in a secure environment. The experiments were conducted on an Intel\textsuperscript{\textregistered} Core\textsuperscript{TM} i7-4700 CPU @ 3.40 GHz $\times$ 8, running a 64-bit Ubuntu 18.04.6 LTS operating system with 16 GB of RAM and 1 TB of disk space. Ubuntu was used as the host OS, with VirtualBox installed for virtualization. A Windows 7 Pro virtual machine was configured with 2 GB of RAM and 32 GB of disk space. The latest version of Cuckoo\cite{cuckoo_sandbox} was installed on the host machine, while the Cuckoo agent was deployed on the virtual machine to facilitate malware analysis. Malware samples were executed in the Windows virtual machine, and activity reports were transmitted to the host Ubuntu system via the Cuckoo agent. Additionally, we employed MetaMe\cite{metame}, a metamorphic code engine, to mutate the samples and analyze their obfuscation behavior. To validate obfuscation, VirusTotal's\cite{virustotal} hash search was utilized for further verification.

\subsection{Dataset}

To create a dataset of rootkit malware, we downloaded 40 rootkit samples from open sources such as MalwareBazaar\cite{malwarebazaar}. These rootkits exhibited properties such as evading detection, privilege escalation, and file manipulation.

Given the limited availability of rootkit samples and to demonstrate obfuscation techniques used by malware authors, we utilized MetaMe\cite{metame} to mutate each malware sample ten times, resulting in 400 mutated rootkits. These samples were divided into:
\begin{itemize}
    \item \textit{310 samples} for analysis and model training
    \item \textit{90 samples} as test data (previously unseen by the model)
\end{itemize}

Additionally, we included \textit{300 benign executables}, consisting of general-purpose applications like browsers and system processes. In total, our dataset comprised \textit{610 executables} for analysis and modeling.

\subsection{Obfuscation Engine: MetaMe}

\begin{figure}[h]
    \centering
    \includegraphics[width=1.0\textwidth]{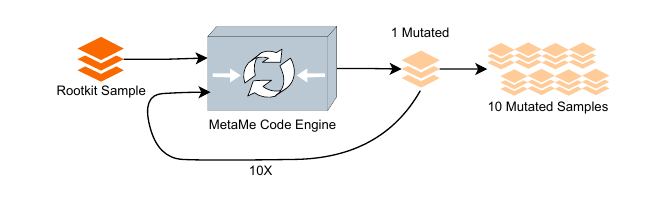}
    \vspace{-5mm}
    \caption{ MetaMe Mutation Architecture}
    \label{fig:dkom}
\end{figure}

MetaMe\cite{metame} is a simple metamorphic code engine that generates logically equivalent versions of executable code. Malware authors leverage metamorphic techniques to evade signature-based detection systems. In our experiment, MetaMe was used to mutate rootkit samples, altering their hash values to remain undetected by traditional antivirus solutions\cite{metame}. MetaMe Architecture Fig~\ref{fig:dkom}
% \vspace{-4mm}

% \vspace{mm}
Each malware sample was passed through MetaMe in a recursive process:
\begin{enumerate}
    \item If $A$ is the original executable, MetaMe generates $A_1$.
    \item $A_1$ is then passed through MetaMe to generate $A_2$.
    \item This process continues iteratively up to $A_{10}$, resulting in ten variations of the original sample ($A_1, A_2, ..., A_{10}$).
\end{enumerate}

These transformed samples were then analyzed to assess their detectability. When tested against VirusTotal\cite{virustotal}, most modified signatures remained undetected, except for a few cases where hash values did not change significantly. 

\subsection{Experimental Methodology}

For our experiment, we prepared \textit{610 executable samples} (310 rootkit malware and 300 benignware).

\subsubsection{Malware and Benignware Execution}

The sandbox environment was set up by creating a virtual network and configuring \textit{Cuckoo}\cite{cuckoo_sandbox} to establish an isolated execution space for malware analysis. Once the sandbox was ready, the collected malware samples were executed, generating detailed \textit{Cuckoo analysis reports}. These reports contained essential behavioral data, including \textit{process creation details, API calls made, files created or modified, network connections established, memory dumps, and Indicators of Compromise (IOCs)}. From these reports, API sequences were extracted alongside their execution timestamps. To enhance the analysis, API calls were grouped into \textit{bigrams} (pairs of API calls) and \textit{trigrams} (triplets of API calls)\cite{malxcap}, incorporating time as a parameter to capture sequence patterns. Only unique API sequences were retained, and \textit{n-gram analysis} was performed to identify patterns indicative of rootkit behavior.

\subsubsection{N-Gram Analysis}

We employed \textit{bi-gram and tri-gram feature engineering} techniques. Given a malware sample $M$ containing an API sequence $m_i = \{a_1, a_2, ..., a_j\}$:
\begin{itemize}
    \item Bigram Sequence: $\{a_1a_2, a_2a_3, ..., a_{j-1}a_j\}$
    \item Trigram Sequence: $\{a_1a_2a_3, ..., a_{j-2}a_{j-1}a_j\}$
\end{itemize}

Using this method, we obtained, \textit{12,007 unique bi-gram sequences} and \textit{68,442 unique tri-gram sequences}. One-hot encoding was applied to prepare datasets for machine learning models.

\subsubsection{Classification Models}
We created two training and testing datasets: a bigram matrix consisting of 610 rows (executables) and 12,008 columns representing API calls, along with one label column, and a trigram matrix with 610 rows and 68,443 dimensions. To classify rootkit behavior, we employed multiple machine learning classifiers from the scikit-learn package, including \textit{Decision Tree (DT), Random Forest (RF), Support Vector Machine (SVM), K-Nearest Neighbors (KNN), Logistic Regression, AdaBoost, XGBoost, and Gradient Boosting}. The dataset was split into a \textit{70:30} ratio for training and testing, ensuring a balanced evaluation of model performance. Additionally, we used the \textit{joblib} library to save trained models, allowing for future classification of previously unseen traces.

\subsection{Classification Results}

In our experiment, we distributed the training and testing data in a 7:3 ratio. Specifically, 70\% of the dataset was used for training the Machine Learning Models, while the remaining 30\% was used for testing the models. In this section, we describe the results obtained using the 30\% testing dataset.

Using this testing dataset, we achieved the best accuracy of \textbf{97.2678\%} for bigram and \textbf{96.1749\%} for trigram API sequence feature matrices using the Random Forest Machine Learning Classifier. Tables~\ref{tab:bigram-results} and ~\ref{tab:trigram-results} present the results of our evaluation of bigram and trigram models, respectively. These tables showcase key performance metrics, including accuracy, precision, recall, and F1-score, for the various machine learning classifiers used to assess model performance.  

% \subsection*{Model Performance on Test Set (Bigram)}

\begin{table}[h!]
\centering
\caption{Model performance on the test set for Bigram API sequences (values in percentages).}
\label{tab:bigram-results}
\vspace{-3pt}
\begin{tabular}{lrrrr}
\toprule
\textbf{Model} & \textbf{Accuracy} & \textbf{Precision } & \textbf{Recall} & \textbf{F1 Score} \\
\midrule
RF            & 97.268  & 97.274  & 97.268  & 97.268  \\
XGBoost       & 96.175  & 96.180  & 96.175  & 96.174  \\
LR            & 96.721  & 96.920  & 96.721  & 96.716  \\
SVM           & 96.175  & 96.443  & 96.175  & 96.167  \\
KNN           & 94.536  & 94.617  & 94.536  & 94.531  \\
GB            & 96.175  & 96.180  & 96.175  & 96.174  \\
DT            & 96.721  & 96.721  & 96.721  & 96.721  \\
AdaBoost      & 95.628  & 95.974  & 95.628  & 95.616  \\
\bottomrule
\end{tabular}
\end{table}

\vspace{-20pt}
\begin{table}[h!]
\centering
\caption{Model performance on the test set for Trigram API sequences (values in percentages).}
\vspace{-3pt}
\begin{tabular}{lrrrr}
\toprule
\textbf{Model} & \textbf{Accuracy} & \textbf{Precision} & \textbf{Recall} & \textbf{F1 Score} \\
\midrule
RF  & 96.175                & 96.443                  & 96.175               & 96.167                 \\
XGBoost        & 94.536                & 95.066                  & 94.536               & 94.514                 \\
LR & 96.175           & 96.443                  & 96.175               & 96.167                 \\
SVM            & 96.175                & 96.443                  & 96.175               & 96.167                 \\
KNN            & 85.246                & 88.566                  & 85.246               & 84.879                 \\
GB & 95.082             & 95.516                  & 95.082               & 95.066                 \\
DT  & 93.989                & 94.625                  & 93.989               & 93.961                 \\
AdaBoost       & 95.082                & 95.516                  & 95.082               & 95.066                 \\
\bottomrule
\end{tabular}
% \caption{Model performance on the test set for bigram API sequences (values in percentages).}
\label{tab:trigram-results}
\end{table}
\vspace{10pt}

\section{Key Feature Importance Analysis}

In our experiment, we utilized two datasets: one for bigram features and another for trigram features. These datasets contain both relevant and irrelevant features, which can impact the detection results. To identify the most significant contributing features, we employed a feature importance for tree-based classifier, which leverages Machine Learning’s feature importance ranking using Gini Impurity Indexing. This method assigns an importance score between \([0,1]\) to each feature.

Tree-based classifiers, such as Decision Trees and Random Forests, select features for splitting based on their ability to reduce Gini Impurity\cite{gini}. Features that contribute to larger reductions in impurity are deemed more important, making them highly relevant for classification. This approach not only aids in feature selection but also helps extract the most informative features that significantly impact detection accuracy. By reducing the dimensionality of the feature matrix, we optimize computational efficiency without compromising detection performance. Our hypothesis is that by selecting only the most relevant features, we can effectively detect rootkit-like behavior with high precision while significantly lowering computational overhead. This ensures a balance between accuracy and efficiency, making our approach fast and scalable.

\subsection{Feature Selection}

To extract the most significant features from bigram and trigram sequences, we utilized Gini impurity-based feature importance indexing. For each feature in the n-gram feature vector, importance scores were computed using tree-based classifiers that rely on Gini impurity to assess the contribution of each feature toward classification performance. These Gini-based importance scores range between 0 and 1, where a higher value indicates a greater role in reducing impurity during the decision-making process. Once computed, we perform the following task based on their Gini importance scores.
\begin{itemize}
    \item For the \textit{bigram dataset}, we had 12,007 features. These were sorted as \(\{ B_1, B_2,\\ ..., B_{12,007} \}\) based on their importance score, ensuring that \( B_i > B_{i+1} \).
    \item Similarly, for the \textit{trigram dataset}, we had 68,442 features, sorted as \(\{ T_1, T_2, \\..., T_{68,442} \}\) where \( T_i > T_{i+1} \).
\end{itemize}

Further for classification results, we divided the sorted features into chunks of 100 and trained the classifier incrementally. Rather than selecting features based on a Gini threshold which would lead to variable chunk sizes and potentially skew the analysis, we chose to divide the sorted features into fixed-size chunks of 100. This fixed-size chunking ensures consistency across experimental runs and allows us to evaluate how model performance scales as more features are included. The process is followed in Algorithm~\ref{alg:feature_selection}:
\vspace{-10pt}
\begin{algorithm}[!ht]
\caption{Incremental Feature Selection and Classification}
\label{alg:feature_selection}
\begin{algorithmic}[1]
\REQUIRE Sorted feature sets for bigram $\{fb_1, fb_2, ..., fb_{12,007}\}$ and trigram $\{ft_1, ft_2, ..., ft_{68,442}\}$, Tree-based classifiers $\{DT, RF, AdaB, XGB, GB\}$
\STATE Initialize chunk size $k \gets 100$
\STATE Initialize maximum F1-score $\text{F1}_{\max} \gets 0$ 
\STATE Initialize best feature count $n_{\max} \gets 0$
\FOR{$i = k$ to $n$ \textbf{step} $k$}
    \STATE Select top $i$ features:
    \STATE \hspace{5mm} For bigram: $\{fb_1, fb_2, ..., fb_i\}$
    \STATE \hspace{5mm} For trigram: $\{ft_1, ft_2, ..., ft_i\}$
    \FOR{each classifier $C$ in $\{DT, RF, AdaB, XGB, GB\}$}
        \STATE Train $C$ with selected features
        \STATE Compute performance metrics (Precision, Recall, F1-score, Accuracy)
        \IF{F1-score $> \text{F1}_{\max}$}
            \STATE Update $\text{F1}_{\max} \gets$ current F1-score
            \STATE Update $n_{\max} \gets i$
        \ENDIF
    \ENDFOR
\ENDFOR
\STATE Output the first $n_{\max}$ features where F1-score is maximized for both bigram and trigram.
\end{algorithmic}
\end{algorithm}

By analyzing the recorded F1-scores, we determined that the \textit{most contributing features correspond to the first chunk where the F1-score reached its maximum}. These features were considered the most relevant for detecting malicious behavior. After reaching the maximum F1-score, additional features often resulted in a decline in model performance. This suggests that while the initial \( n \times 100 \) features contributed positively, subsequent features may have introduced noise or irrelevant information, thereby reducing accuracy.

\begin{figure*}[!h]
    \centering
    \includegraphics[width=1.0\textwidth]{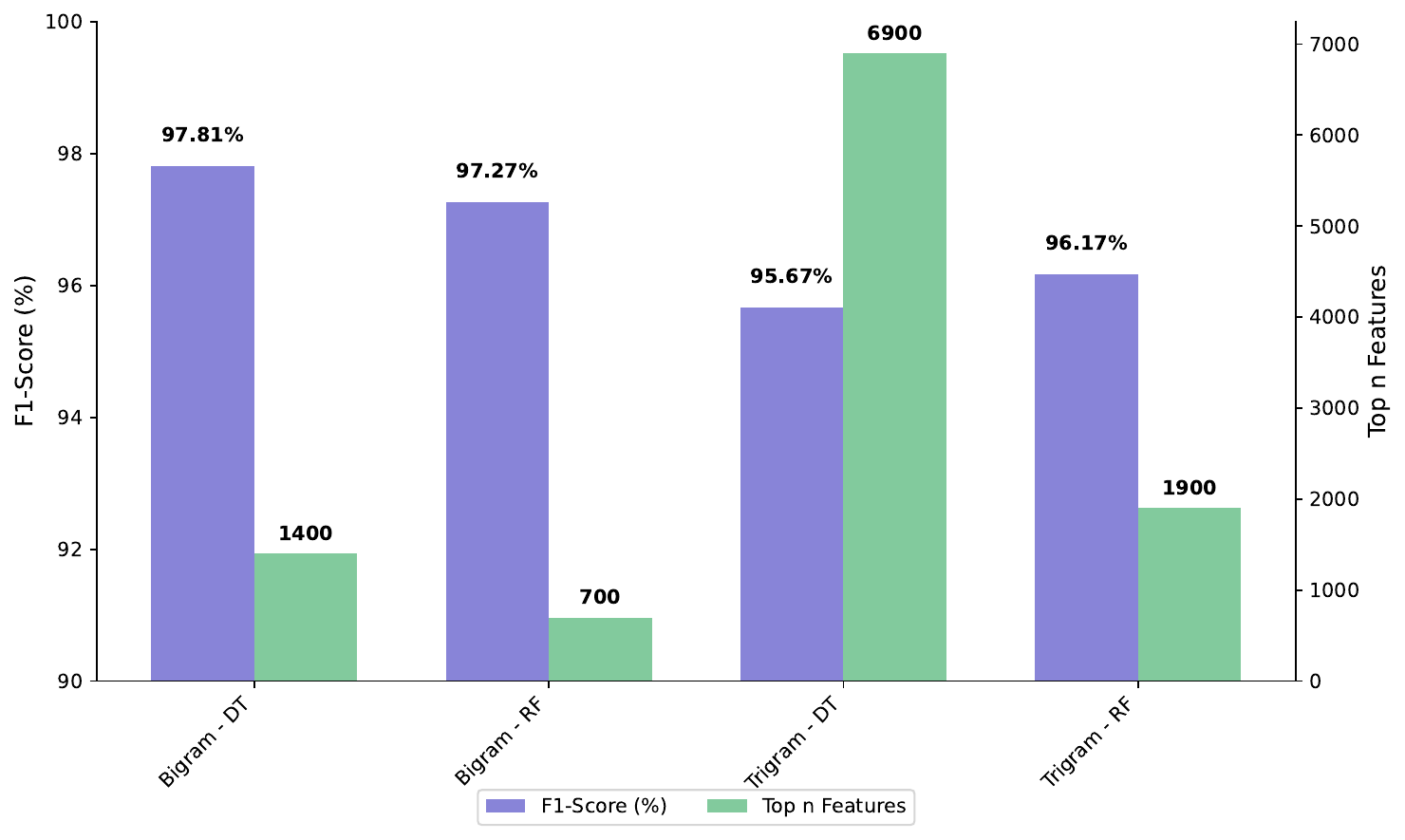}
   \caption{Model performance comparison for Top n-feature where F1-score is maximum}
    \label{fig:Model_Performance}
\end{figure*}

Interestingly, we observed fluctuations in results when using Decision Tree and Random Forest classifiers. However, other models such as \textit{XGBoost, AdaBoost, and Gradient Boosting} showed more stable results across all feature chunks. Therefore, \textit{Decision Tree and Random Forest were chosen for further analysis} to identify features that exhibit rootkit-like behavior. In Fig.\ref{fig:Model_Performance}, we can see the bar graph where the Decision Tree classifier reaches the maximum F1-score at Top \( n \times 100 \) features where n is 14, shorted using Gini Impurity Index. Similarly, we considering the Decision Tree classifier for Trigram with Top \( n \times 100 \) features where n is 69 and 19 in the case of Random Forest, where the F1-score is nearly same both the case. 

\vspace{-10mm}
\subsection{Identifying the Most Contributing Features}
\vspace{-1mm}
To pinpoint the most relevant features, we identified the intersection of the top \( n \times 100 \) features from both Decision Tree and Random Forest classifiers where the F1-score is max. The overlapping features were considered the most informative for rootkit detection. 
Let us consider the set {A, B, C, D} for the selected feature sets for bigram and trigram classifiers be defined as follows:
\vspace{-3 pt}
\begin{align*}
    A &= \{ fb_1^{DT}, fb_2^{DT}, \dots, fb_{1,400}^{DT} \}   \\
    B &= \{ fb_1^{RF}, fb_2^{RF}, \dots, fb_{700}^{RF} \}   \\
    C &= \{ ft_1^{DT}, ft_2^{DT}, \dots, ft_{6,900}^{DT} \}   \\
    D &= \{ ft_1^{RF}, ft_2^{RF}, \dots, ft_{1,900}^{RF} \}  
\end{align*}
% \vspace{-3pt}
To find features that are important across models, we look for overlaps in their top-ranked features. This is done by calculating the intersection of the feature sets from each model. In our case, we compared the most significant features chosen by both the Random Forest and the Decision Tree classifiers, using both bigram and trigram feature vectors. If a feature appears in the lists of both classifiers, it means that feature is consistently recognized as important. These overlapping features are valuable because they show strong and stable influence on the model classification. 
% \vspace{-3 pt}
\begin{align*}
    |A \cap B| &= 76  \\
    |C \cap D| &= 483
\end{align*}
\vspace{-5 pt}

For bigram we have found 76 common features from the Decision Tree and Random Forest classifier where we have seen the first F1-score max. Similarly, for trigram we have found 483 common features which are really contributing. Our approach demonstrates high reduction of the feature matrix dimension like for bigram, we had 12,007 features contains both relevant and irrelevant features, now we have 76 features which are more relevant and really contributing towards results for rootkit detection, and for the trigram, we had 68,442 features now we have 483 most relevant features. With our approach, in both bigram and trigram we have seen a heavy reduction of dimension from higher dimension to lower dimension. We have further trained the machine learning classifier based on these lower dimension i.e. considering only most relevant features. While have tested this hypothesis using the 30\% testing data that was split from feature matrix for model testing for the higher dimension data. We have successfully achieved the accuracy of \textbf{96.72\%} for bigram and \textbf{97.81\%} for trigram using the Random Forest classifier. These results demonstrates our hypothesis for considering the most relevant features meanwhile decrease featured dimension drastically. For bigram and for trigram sequences, a few listed features  are listed in Table~\ref{tab:bigram features} for Bigram Sequences and Table~\ref{tab:trigram feature}  for Trigram Sequences along with Justification.

\begin{table}[h!]
\centering
\caption{Top Common Features for Bigram Features.}
\label{tab:bigram features}
\begin{tabular}{|p{4cm}|p{8cm}|}
\hline
\textbf{API Sequence} & \textbf{Justification} \\
\hline \vspace{3pt}
(GetTempPathW, DeviceIoControl) & DeviceIoControl, to interact with device drivers, especially in the temp directory, may indicate attempt exploit vulnerabilities in drivers or install rootkits. \\
\hline \vspace{3pt}
(GetTempPathW, NtCreateFile) & Temp directory/file creation in low-level sys call (ntoskrnl), could be used for staging malicious payload. This rootkit behavior can modify critical system files to facilitate stealth or control. \\
\hline \vspace{3pt}
(GetSystemTimeAsFileTime, NtQueryKey) & Querying registry keys after system time retrieval may indicate malicious configuration checks. Rootkit behavior for creating unique payloads or registry based on timestamp. \\
\hline \vspace{3pt}
(GetSystemDirectoryW, RegOpenKeyExW) & Querying the system directory and accessing the registry are activities that, when combined, could indicate an attempt to tamper with system configurations. \\
\hline\vspace{3pt}
(HttpQueryInfoA, RegOpenKeyExW) & HTTP query combined with registry access suggests network-based configuration tampering. Allows rootkit to dynamically adapt to instructions from remote servers for long term malware activity. \\
\hline
\end{tabular}
\vspace{3pt}
% \caption{Top Common Features for Bigram Features.}
\end{table}

\begin{table}[h!]
\centering
\caption{Top Common Features for Trigram Features.}

\begin{tabular}{|p{4cm}|p{8cm}|}
\hline
\label{tab:trigram feature}
\textbf{API Sequence} & \textbf{Justification} \\
\hline\vspace{3pt}
(IsDebuggerPresent, DeviceIoControl, NtQuerySystemInformation) & Rootkits often use DeviceIoControl to interact with stealth-enabling kernel components. Malware and rootkits often use NtQuerySystemInformation to manipulate or hide processes or modules. \\
\hline\vspace{3pt}
(IsDebuggerPresent, CreateDirectoryW, DeviceIoControl) & The sequence has rootkit-like potential if the APIs are being used to evade detection, hide malicious files, or communicate with system-level drivers (IRPs), potentially as part of privilege escalation and maintaining persistence. \\
\hline\vspace{3pt}
(IsDebuggerPresent, CreateThread, NtCreateFile) & Checking the debugging environment, and CreateThread could be used for spawning threads to perform malicious actions. NtCreateFile could be used to create or manipulate files, including hidden or system files. \\
\hline\vspace{3pt}
(LdrGetProcedureAddress, NtAllocateVirtualMemory, NtProtectVirtualMemory) & Both NtAllocateVirtualMemory and NtProtectVirtualMemory are associated with memory manipulation, commonly used by rootkits to manipulate or hide malicious code in memory. \\
\hline\vspace{3pt}
(IsDebuggerPresent, GetSystemDirectoryW, NtAllocateVirtualMemory) & Debugger detection, system directory access, and memory allocation all point toward a potential rootkit attempting to hide itself in system memory or files. Rootkit behavior related to memory manipulation and evasion. \\
\hline
\end{tabular}
\vspace{-3pt}
% \caption{Top Common Features for Trigram Features.}
% \vspace{30pt}

\end{table}

\section{Model Testing}

During our experiment, we utilized \textit{MetaMe} to mutate malware samples, ensuring that 90 samples remained undetected by the model. These rootkit samples were intentionally excluded from the training phase to evaluate our model’s ability to detect unknown threats.
\vspace{15mm}
\begin{figure}[!h]
    \centering
    \includegraphics[width=1.0\textwidth]{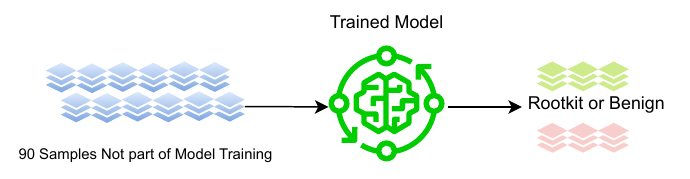}
    \caption{Model Prediction for Unseen Samples}
    \label{fig:Model_Prediction}
\end{figure}
\vspace{7mm}
To evaluate the effectiveness of our approach, we extracted API sequence features from unknown rootkit samples and tested them using our trained machine learning classifier. We conducted experiments on both high-dimensional feature matrices (containing both relevant and irrelevant features) and optimized lower-dimensional feature sets (comprising only the most significant features). Using a dataset of 90 previously unseen rootkit samples.

For the higher-dimensional feature matrix, where both relevant and irrelevant features were included, the Random Forest classifier correctly identified 77 out of 90 samples as rootkits for bigram sequences and 70 out of 90 samples for trigram sequences. However, when using the lower-dimensional feature matrix, which retained only the most relevant features, the classifier performed better by correctly identifying 80 out of 90 samples for bigrams as a rootkit and achieving a perfect detection rate of 90 out of 90 for trigrams.

These findings validate our hypothesis that analyzing API call sequences enhances rootkit detection. The improved detection rates with optimized feature selection confirm that reducing dimensionality enhances classification accuracy and minimizes computational overhead. Additionally, the high detection rate for previously unseen rootkits demonstrates the model’s strong generalization ability, making it a viable approach for detecting sophisticated and evolving malware threats in real-world environments.

\section{Limitations and Future Work}

Future research can focus on expanding the rootkit database, which is currently limited in scope via publicly available database. Institutions with access to larger and more diverse datasets could facilitate the development of a more robust model by contributing additional samples of known rootkits and rootkit-like behaviors.
Further, we can extend our approach to quadgrams or more consecutive sequences. However, incorporating quadgrams would significantly increase the feature matrix, potentially up to 10 times, leading to higher computational overhead. Future work should explore optimized feature selection and dimensionality reduction techniques to balance performance with efficiency when scaling to higher-order N-grams. 
Meanwhile, our approach has demonstrated potential in detecting rootkits or rootkit-like behavior. It may flag off generic processes as rootkits, processes such as privilege escalation, that may resemble rootkit-like activity. But this drawback can be overcome by training on a broader rootkit dataset and incorporating higher-order n-gram features.

\section{Conclusion}

SeqShield presents a robust approach for extracting API sequences and training machine learning classifiers to detect rootkits. Our method achieved an impressive 97.27\% accuracy using a bigram feature matrix and 96.17\% accuracy with a trigram feature matrix, both evaluated with the Random Forest Classifier. Additionally, we demonstrated how rootkits leverage obfuscation techniques to evade traditional signature-based IDS.
By leveraging the Gini impurity index, we effectively reduced the feature space, retaining only the most impactful features while eliminating irrelevant ones. This reduction significantly decreased computational overhead and reduced the dimensionality of the featured matrix without compromising model accuracy. Furthermore, we validated SeqShield’s detection capabilities by analyzing previously unseen rootkit traces, confirming its effectiveness in identifying modern and evasive kernel-level rootkits.

\bibliographystyle{splncs04}
\bibliography{reference}

\end{document}